\documentclass[twocolumn,runningheads]{svjour2}
\smartqed  
\usepackage{graphicx}
\journalname{Astrophysics and Space Science}
\begin{document}

\title{GRBs search results with the ARGO-YBJ experiment operated in Scaler Mode}
\author{G. Di Sciascio and T. Di Girolamo\\
 for the ARGO-YBJ Collaboration}

\institute{G. Di Sciascio \at
              INFN, sez. di Napoli \\
              Tel.: +39-81-676453\\
              Fax: +39-81-676346\\
              \email{giuseppe.disciascio@na.infn.it}           
}

\date{Received: date / Accepted: date}

\maketitle

\begin{abstract}
The ARGO-YBJ experiment is almost completely installed at the
YangBaJing Cosmic Ray Laboratory (4300 m a.s.l., Tibet, P.R.
China). The lower energy limit of the detector (E $\sim$ 1 GeV) is
reached with the scaler mode, i.e., recording the single particle
rate at fixed time intervals. In this technique, due to its high
altitude location and large area ($\sim$ 6700 m$^2$), this
experiment is the most sensitive among all present and past
ground-based detectors. In the energy range under investigation,
signals due to local (e.g. solar GLEs) and cosmological (e.g.
GRBs) phenomena are expected as significant enhancements of the
counting rate over the background. Results on the search for GRBs
in coincidence with satellite detections are presented.
\end{abstract}

\section{Introduction}
\label{intro}

Gamma Ray Bursts (GRBs) have been deeply studied in the keV -- MeV
energy range, but only little information in the GeV range has
been provided in the past decade by EGRET measurements. Only 3
bursts have been detected at energies $>$1 GeV, with one photon
reaching 18 GeV \cite{egret}. The study of the high energy
emission of GRBs could provide extremely useful data able to
constrain the emission models and the value of the ambient
parameters. At these high energies the detection from space is
hampered by the very low fluxes, requiring large collection areas.
From ground the last generation Cherenkov telescope MAGIC
\cite{magic}, designed also to point at the detected GRB direction
in a very short time, is still making efforts to lower the
threshold at energies $\leq$100 GeV. Its duty cycle and field of
view are however very small: a wide field detector, able to cover
simultaneously and continuously a significant ($\sim$1 steradian)
fraction of the sky, is thus necessary. With such a detector the
Milagro Collaboration reported evidence for TeV emission from GRB
970417a with a significance slightly greater than 3$\sigma$
\cite{milagrito}. The study of the high energy spectrum of GRBs is
perhaps the strongest motivation for an all-sky VHE detector. The
study of transient phenomena can be successfully performed at
energies down to 1 GeV by air showers arrays working in "single
particle mode" \cite{vernetto}, i.e., counting all the particles
hitting the individual detectors during fixed time intervals. The
observation of an excess in coincidence with a GRB detected by
satellites would be an unambiguous signature of the nature of the
signal. Both the sensitivity and the energy threshold improve with
larger detection areas and higher observation levels, making air
shower detectors at very high altitude the most suitable.

The ARGO-YBJ experiment, located at the YangBaJing Cosmic Ray
Laboratory (4300 m a.s.l.) with a detection area of $\sim$ 6700
$m^2$, is an air shower array exploiting the full coverage
approach at very high altitude, with the aim of studying the cosmic
radiation with a low energy threshold. In this paper we present results
on the search for GRBs in coincidence with satellite detections
performed in the December 2004 - May 2006 period with the ARGO-YBJ
experiment.

\section{The detector}

The ARGO-YBJ detector is constituted by a single layer of Resistive Plate
Chambers (RPCs) with $\sim$93$\%$ of active area. This
carpet has a modular structure, the basic module being a cluster
(5.7$\times$7.6 m$^2$), divided into 12 RPCs (2.8$\times$1.25
m$^2$ each). Each chamber is read by 80 strips of 67.5$\times$618
mm$^2$, logically organized in 10 independent pads of
55.6$\times$61.8 cm$^2$ which are individually acquired and
represent the high granularity pixel of the detector
\cite{nim_argo}. The carpet is composed by 154 clusters for a
total surface of $\sim$6700 m$^2$.

The detector is connected to two different DAQ systems, which work
independently: in shower mode, for each event which fulfill the
trigger conditions the position and time of each detected particle
is recorded, allowing the reconstruction of the lateral
distribution and of the arrival direction \cite{ang_icrc05}; in
scaler mode, where there is no trigger, the counting rate of each
cluster is measured every 0.5 s, with no measurement of the space
distribution and arrival direction of the detected particles. In
the scaler mode DAQ, for each cluster the signal coming from the
120 pads is added up and put in coincidence in a narrow time
window (150 ns), giving the rate of counts $\ge$1, $\ge$2, $\ge$3,
$\ge$4, read by four independent scaler channels. The
corresponding measured rates are, respectively, $\sim$ 40 kHz,
$\sim$ 2 kHz, $\sim$ 300 Hz and $\sim$ 120 Hz for each cluster.
The counting rates for a given multiplicity are then obtained with
the relation n$_i$ = n$_{\geq i}$ - n$_{\geq i+1}$ for i = 1, 2,
3. The use of four different scalers may give an indication of the
source spectrum in case of signal detection. In order to correctly
handle the data, it is very important to evaluate the response to
particles hitting the detector. For scaler mode operations, the
most important effect is the strip cross-talk, i.e., the
probability of having more than one strip fired by a single
particle, giving fake coincident counts. Due to the front-end
logic, this can happen only for strips belonging to different
pads, since the maximum number of counts for each pad is 1
independently of the number of particles hitting simultaneously
the pad. An analytical calculation based on the measured
"occupancy", i.e., the mean number of strips fired by 1 particle,
has been made and checked experimentally.

\begin{figure}
\centering
  \includegraphics[height=.30\textheight]{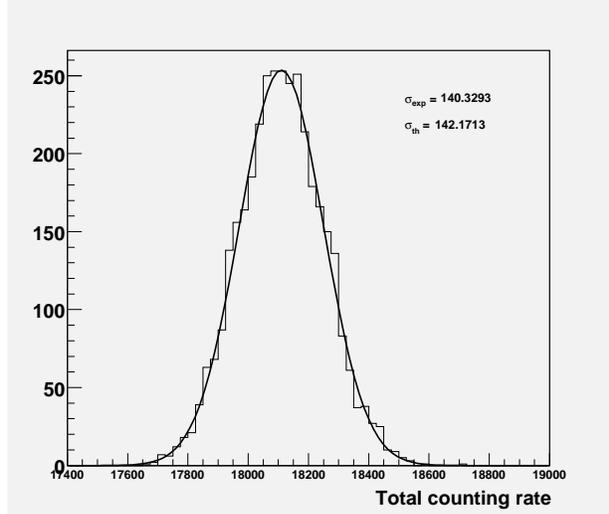}
\caption{Total counting rate summed over the 4 multiplicity
channels for a typical cluster.}
\label{fig:csum}       
\end{figure}

From the experimental point of view, it is important to take into
account the background counting rate variations due to changes in
environmental parameters such as the atmospheric temperature and
pressure (which modify the shower development in the atmosphere)
and the detector temperature (instrumental effect). More
troublesome are other possible instrumental effects, such as the
electronic noise, that could simulate narrow signals in time,
producing spurious increases in the background rate. Working in
single particle mode requires very stable detectors, and a very
careful and continuous monitoring of the experimental conditions.
By comparing the counting rate of the single detectors and
requiring simultaneous and consistent variations in all of them,
it is possible to identify and reject most of the fake excesses
due to instrumental effects. Short variations in the single
particle counting rate have been measured in coincidence with
strong thunderstorms and have been ascribed to the effects of
atmospheric electric fields on the secondary particles flux
\cite{thunderst}. The static electric field is measured on the
roof of the ARGO-YBJ building with an EFM100 Boltek atmospheric
field monitor. Anyway, we note that the occurrences of these
events are very rare and even in this case the observed time
scales ($\sim$10 - 15 minutes) are longer than the typical GRB
duration. The study of the counting distribution for each cluster
is important in order to monitor the stability of the detector and
its statistical (Poissonian) behaviour. In Fig. \ref{fig:csum} the
total counting rate of a typical cluster, added up on the 4
multiplicity channels during a period of 30 minutes, follows a
Poissonian distribution with a $\sigma^2$ given by:
$$\sigma^2 (C_{tot}) = \sigma^2 (C_1) + 4\cdot\sigma^2 (C_2) + 9\cdot\sigma^2 (C_3) + 16\cdot\sigma^2 (C_4)$$
\section{Determination of the detector effective areas}

\begin{figure*}
\begin{minipage}[t]{.47\linewidth}
  \begin{center}
   \includegraphics[height=.30\textheight]{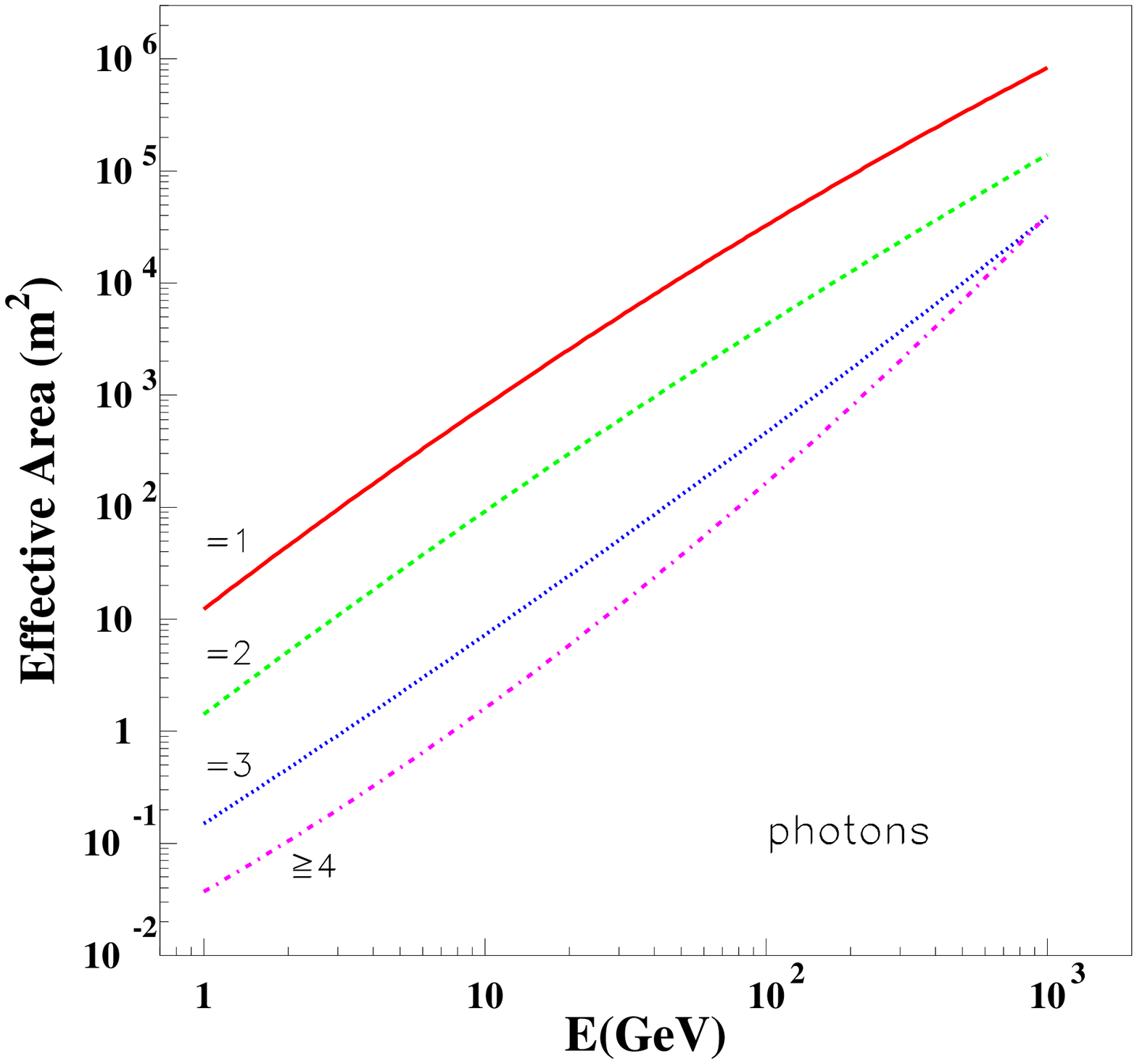}
  \end{center}
\end{minipage}\hfill
\begin{minipage}[t]{.47\linewidth}
  \begin{center}
  \includegraphics[height=.30\textheight]{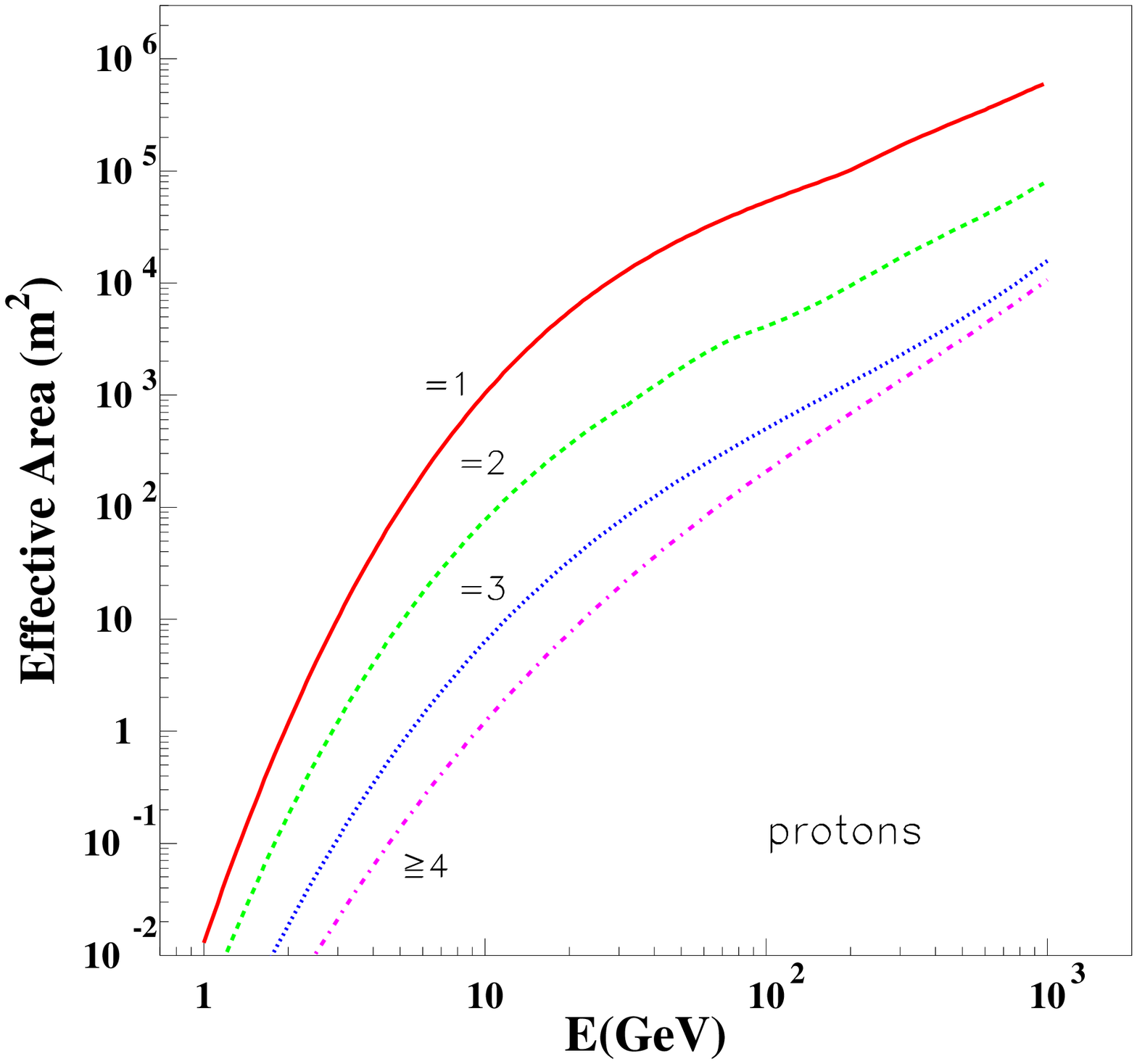}
  \end{center}
\end{minipage}\hfill
  \caption{Effective areas versus energy for primary photons (left plot) and protons
(right plot) with zenith angle $\theta = 20^{\circ}$. The curves
refer to different multiplicity channels: n = 1, n = 2, n = 3 and
n $\geq$ 4.}
\label{fig:aeff_20g}
\end{figure*}

In order to study the detector response to Extensive Air Showers
(EASs), a detailed MC simulation has been carried out for both
protons and photons with fixed energies in the range 1 GeV - 1 TeV
and zenith angles $\theta$ = 0$^{\circ}$, 10$^{\circ}$,
20$^{\circ}$, 30$^{\circ}$, 40$^{\circ}$. The CORSIKA/QGSJet code
6.204 \cite{corsika} has been used with a full electromagnetic
component development down to E$_{thr}$= 0.05 MeV for both
electrons and photons and 50 MeV for muons and hadrons. A detailed
description of the detector has been carried out to correctly
simulate the {\emph "cluster size"}, i.e., the correlation between
the number of particles hitting the detector and the number of
signals generated in the different multiplicity channels. Since
the actual efficiency depends essentially on the shower particle
lateral distribution, a huge quantity of showers must be simulated
over a very large area to completely contain it. To save the
computing time the shower sampling can be performed by means of
the {\em "reciprocity technique"} \cite{battist}. The sampling
area A$_S$ ($\sim$ 5000$\times$5000 m$^2$) is uniformly filled
with replicas of the same ARGO-YBJ carpet, one adjacent to the
other. Following the reciprocity concept, we sample the shower
axis only over the area covered by the carpet located at the
center of the array, with the prescription of considering the
response of all the detector replicas. On an event-by-event basis
we calculate the number of clusters which contain more than 1, 2,
3, 4 fired pads, summed on the entire grid. Fig.\ref{fig:aeff_20g}
shows the effective areas for primary photons and protons with
zenith angle $\theta = 20^{\circ}$ in the four multiplicity
channels for the complete ARGO-YBJ detector constituted by 154
clusters ($\sim$6700 m$^2$ sensitive area). We note that for a
multiplicity $n=1$ the detector sensitivity does not depend on its
geometrical features, like the area of the single counters or
their relative positions, but only on the total sensitive area
\cite{vernetto}. Therefore, the effective areas for any carpet
dimension can be scaled from the plotted values.
%
%
The effective areas for primary protons are then convoluted with
the following spectrum: $dN_p /dE \propto E^{-\Gamma}$ with
$\Gamma =2.7$ \cite{gaisser} and taking into account the local
geomagnetic cutoff \cite{storini}. The resulting counting rates,
considering an opening angle of 60 degrees around the zenith, are
the following: 21 kHz for n = 1, 1.7 kHz for n = 2, 180 Hz for n =
3 and 80 Hz for n $\geq$ 4. Comparison with the measured rates,
i.e., 38 kHz for n = 1, 1.7 kHz for n = 2, 180 Hz for n = 3 and
120 Hz for n $\geq$ 4, shows that the values obtained by our
simulations are lower in the multiplicity channels n = 1 and n
$\geq$ 4. The discrepancy for n = 1 is expected because of dark
counting and natural radioactivity. Since from both of them we
expect mostly single counts, these effects are expected to
influence only the $\ge$1 scaler channel.

\section{Data Analysis and Results}

The search for emission from GRBs started with the first GRB
detection by the Swift satellite on December 17, 2004, when only
16 clusters ($\sim$693 m$^2$ of sensitive area) out of the total
154 were in data taking. Up to May 2006, 28 GRBs detected by
satellites were within the ARGO-YBJ field of view (for this
search, $\theta \leq 40^{\circ}$). Because of detector
installation and debugging operations, the duty cycle of data
taking has been reduced and reliable data are available only for
16 of these GRBs (see Table \ref{tab:grb}).

For every GRB, the number of counts $N$, recorded in each of the
four multiplicity channels during the duration time $T90$ measured
by the satellites, is compared with the number $B$ expected from
the background (obtained from the average counting rate in $\pm
10\cdot T90$ around the burst). The difference $N-B$ in units of
standard deviations, i.e., $(N-B)/\sqrt{B+B/20}$, gives the
statistical significance $n{_\sigma}$ of the excess, which we report in
column 8 of Table \ref{tab:grb} for $n=1$.

The data analysis of 3 GRBs (GRB051114, GRB060105 and GRB060510A)
gives 2.8, 3.6 and 3.7 as the statistical significance of the
signal, respectively. Taking into account that we considered a
sample of 16 GRBs, these values correspond to a post-trial
probability P($>1.7\sigma$), P($>2.8\sigma$) and P($>2.9\sigma$),
respectively, of being a background fluctuation. As a consequence,
no convincing excess in the scaler counts was observed in the
duration time measured by the satellites.
%
%
Therefore $3\sigma$ upper limits to the fluence of these events
were calculated in the 1 -- 100 GeV energy range using the spectral
indices determined at lower energies by satellites. Our results are
reported in the last column of Table \ref{tab:grb}. For those GRBs
whose redshift has been also determined, the upper limit was
calculated including a model for $\gamma \gamma$ absorption by
the Extragalactic Background Light (EBL) \cite{absorpt} and the
corresponding values printed in bold. For the other GRBs z
= 0 was assumed (below 300 GeV the $\gamma \gamma$ absorption is
almost negligible for z$<$0.2).

%
\begin{table*}[t]
\begin{center}
\caption{List of GRBs in the field of view ($\theta\leq 40^{\circ}$) of
ARGO-YBJ (Dec. 2004 - May 2006), with preliminary fluence upper limits.}
\label{tab:grb} \vspace{0.3cm}
    \begin{tabular}[h]{llccccccc}
    \hline\noalign{\smallskip}
    \tableheadseprule\noalign{\smallskip}
    GRB & Sat. & T90/Dur. & $\theta^{\ast}$ & Redshift & Spectral & Carpet       & $n_{\sigma}^{\S}$ & UL${^\dag}$ \\
        &          & (s)  & (deg)           &          & Index    & Area (m$^2$) &               & (Fluence)   \\[3pt]
    \hline
 041228 & Swift & 62    & 28.1 & --    & 1.56 & 693  & -1.3  & 3.3$\cdot$10$^{-4}$ \\
 050408 & HETE  & 15    & 20.4 & 1.24  & 1.98 & 1820 & -2.2  & {\bf 9.6$\cdot$10$^{-5}$} \\
050509A & Swift & 12    & 34.0 & --    & 2.1  & 1820 & 0.29  & 1.6$\cdot$10$^{-4}$ \\
 050528 & Swift & 11    & 37.8 & --    & 2.3  & 1820 & -0.012 & 6.5$\cdot$10$^{-4}$\\
 050802 & Swift & 20    & 22.5 & 1.71  & 1.55 & 1820 & 0.74  & {\bf 1.0$\cdot$10$^{-4}$} \\
051105A & Swift & 0.3   & 28.5 & --    & 1.33 & 3379 & 0.90  & 1.4$\cdot$10$^{-5}$ \\
 051114 & Swift & 2     & 32.8 & --    & 1.22 & 3379 & 2.8   & 1.9$\cdot$10$^{-5}$ \\
 051227 & Swift & 8     & 22.8 & --    & 1.31 & 3379 & 0.93  & 2.5$\cdot$10$^{-5}$ \\
 060105 & Swift & 55    & 16.3 & --    & 1.11 & 3379 & 3.6   & 5.9$\cdot$10$^{-5}$ \\
 060111 & Swift & 13    & 10.8 & --    & 1.63 & 3379 & 0.82  & 2.5$\cdot$10$^{-5}$ \\
 060115 & Swift & 142   & 16.6 & 3.53  & 1.76 & 4505 & -2.2  & {\bf 2.3$\cdot$10$^{-4}$} \\
 060421 & Swift & 11    & 39.3 & --    & 1.53 & 4505 & -0.46 & 1.6$\cdot$10$^{-4}$ \\
 060424 & Swift & 37    & 6.7  & --    & 1.72 & 4505 & 1.9   & 4.1$\cdot$10$^{-5}$ \\
 060427 & Swift & 64    & 32.6 & --    & 1.87 & 4505 & -1.8  & 1.8$\cdot$10$^{-4}$ \\
060510A & Swift & 21    & 37.4 & --    & 1.55 & 4505 & 3.7   & 2.3$\cdot$10$^{-4}$ \\
 060526 & Swift & 14    & 31.7 & 3.21  & 1.66 & 4505 & 0.75  & {\bf 1.2$\cdot$10$^{-4}$} \\
    \noalign{\smallskip}\hline
    \end{tabular}
\end{center}
$\ast$ Zenith angle. \\
$\S$ Significance of the signal for the single event.\\
$\dag$ Upper Limit on the fluence (1 -- 100 GeV) in erg cm$^{-2}$.
The numbers in bold take into account absorption by the EBL.\\
\end{table*}

\section{Conclusions}

A search for VHE emission from GRBs has been performed with an
increasing detector area of the ARGO-YBJ experiment. A total of 16
satellite-triggered GRBs in the field of view ($\theta\leq 40^{\circ}$) of
ARGO-YBJ in the Dec. 2004 - May 2006 period has been analyzed.
No significant emission was detected and typical fluence
upper limits of $\approx 10^{-4}$ erg cm$^{-2}$
in the 1 -- 100 GeV energy range were obtained using
the measured counting rates and GRB parameters determined by the
satellite observations.
We expect to increase the sensitivity by a factor $\sim$2
converting the secondary photons with a 0.5 cm thick layer of lead.


\begin{thebibliography}{00}

\bibitem{egret}
Catelli, J.R. et al.: In: C.A. Meegan (ed.), AIP Conf. Proc. No.
428, p. 309 (AIP, New York, 1998).
\bibitem{magic}
Albert, J. et al.: ApJ Lett. {\bf 641} L9 (2006).
\bibitem{milagrito}
Atkins, R. et al.: ApJ Lett. {\bf 553} L119 (2000).
\bibitem{vernetto}
Vernetto, S.: Astrop. Phys. {\bf 13} 75 (2000).
\bibitem{nim_argo}
Aielli, G. et al.: NIM {\bf A562} 92 (2006).
\bibitem{ang_icrc05}
Di Sciascio, G. et al.: 29th ICRC, Pune, {\bf 2}, 33-36 (2005).
\bibitem{thunderst}
Aglietta, M. et al.: 26th ICRC , Salt Lake City, {\bf 7}, 351-354
(1999).
\bibitem{corsika}
Heck, D. et al.: Report {\bf FZKA 6019} Forschungszentrum
Karlsruhe, 1998.
\bibitem{battist}
Battistoni, G. et al.: Astrop. Phys. {\bf 7} 101 (1997).
\bibitem{gaisser}
Gaisser, T.K., Honda M.: Ann. Rev. Nucl. Part.  Sci. {\bf 52} 153
(2002).
\bibitem{storini}
Storini M., Smart D.F., Shea M.A.: In: 27th ICRC, Hamburg, {\bf
10} 4106-4109 (2001).
\bibitem{absorpt}
Kneiske, T.M. et al.: A$\&$A {\bf 413} 807 (2004).


\end{thebibliography}
\end{document}